\documentclass[12pt]{iopart}

\usepackage{amssymb}
\usepackage{latexsym}
\usepackage[dvips]{color,graphicx}
\begin{document}

\title[Quasi-condensates out of Mott-insulators]
{Coherent matter waves emerging from Mott-insulators}

\author{K.\ Rodriguez$^1$, S.R.\ Manmana$^{1,2}$, M.\ Rigol$^3$, 
R.M.\ Noack$^2$, and A.\ Muramatsu$^1$}

\address{
$^1$Institut f\"ur Theoretische Physik III, 
Universit\"at Stuttgart, 70550 Stuttgart, Germany \\
$^2$AG Vielteilchennumerik, Fachbereich Physik, 
Philipps-Universit\"at Marburg, D-35032 Marburg, Germany \\
$^3$Physics Department, University of California, Davis,
CA 95616, USA
}
\begin{abstract}
We study the formation of (quasi-)coherent matter waves emerging from a
Mott insulator for strongly interacting bosons on a
one-dimensional lattice. 
It has been shown previously that a quasi-condensate emerges at momentum
$k_{\rm cond} = \pi/2a$, where $a$ is the lattice constant, in the
limit of infinitely strong repulsion (hard-core bosons).
Here we show that this phenomenon persists for all values of the
repulsive interaction that lead to a
Mott insulator at a commensurate filling. 
The non-equilibrium dynamics of hard-core bosons is treated
exactly by means of a Jordan-Wigner transformation, and the generic case
is studied using a time-dependent density matrix renormalization group
technique. 
Different methods for controlling the emerging matter wave are 
discussed.
\end{abstract}

\pacs{03.75.Kk, 03.75.Lm, 03.75.Pp,05.30.Jp}
\maketitle

\section{Introduction}
Cold atoms in optical lattices currently constitute one of the most
flexible and, hence, most promising ways to investigate areas of
many-particle physics for which no established knowledge prevails. 
One such area is that of strongly correlated systems, for which
neither the ground-state nor excited states can be easily described
and for which no generic analytic tools with predictive power are
presently available. 
The situation is even more difficult for non-equilibrium
processes, where, in general, a full knowledge of the eigenstates of
the system is needed in order to properly describe its dynamics out of
equilibrium. 

Recent experimental work has succeeded in producing one of the
hallmarks of strong correlations, namely, a Mott insulator
\cite{greiner02,stoferle04}. 
Remarkably, this was achieved with bosons, 
a situation not easily
encountered in condensed matter physics.
Even the limit of very strong interactions, where the bosons can be
considered to be impenetrable particles (hard-core bosons or
a Tonks-Girardeau gas \cite{girardeau60}) could be reached experimentally
\cite{paredes04,kinoshita04}.  
Such a limit is attainable in elongated traps for large positive
three-dimensional scattering lengths, at low densities, or with very
strong transversal confinement \cite{olshanii98,petrov00,dunjko01}. 

On the other hand, the study of the non-equilibrium dynamics of
quantum gases was instrumental for understanding their
properties \cite{dalfovo99}.  
More recently, a controlled study of the evolution of one-dimensional
Bose gases prepared initially out of equilibrium was performed
\cite{kinoshita06}, opening up the possibility of investigating the role
of integrability in the relaxation dynamics of a many-body system 
\cite{rigol06}. In the case of hard-core bosons in one dimension, 
the evolution of a system  with an arbitrary initial state can be 
described theoretically in an exact way \cite{rigol04c,rigol05a,rigol05c} 
via an exact mapping onto free fermions, 
the Jordan-Wigner transformation \cite{jordan28}.

We concentrate here on the out-of-equilibrium evolution of a
one-dimensional Bose gas with strong interactions whose
initial state is a Mott insulator. 
It was previously shown that, in the hard-core limit, an initial Fock
state develops quasi-long-range correlations when the bosons are
allowed to evolve freely on a lattice \cite{rigol04c,rigol05c}. 
In particular, the momentum distribution function 
$n_k \equiv \langle n_k\rangle $ develops
sharp peaks at momenta $k=\pm \pi/2a$, where $a$ is the lattice
constant. 
An examination of the one-particle density matrix shows that, after the
formation of the peaks in $n_k$, it decays as $1/\sqrt{x}$ at
long distances, as it does for hard-core bosons in equilibrium
\cite{rigol04b,rigol05b}, demonstrating that, in fact, the peaks in
$n_k$ signal the emergence of quasi-coherence at a finite wavevector. 
A detailed picture of the (quasi-)coherent part is obtained
by examining the lowest natural orbital (NO), i.e., the eigenvector
of the one-particle density matrix corresponding to the largest
eigenvalue. 
Once the peaks in $n_k$ form, the NO evolve at a constant velocity
$v_{NO} = \pm 2 a t/\hbar$, where $t$ is the nearest neighbor hopping
amplitude, without appreciable change in their form.
These are the maximal group velocities on a lattice with a dispersion
$\epsilon_k = -2t \cos ka$. 
The process of formation of the quasi-condensate is also characterized
by a power law. 
The population of the quasi-condensate increases in a universal way as
$\sim 1.38 \sqrt{t \tau/\hbar}$, as a function of the evolution time
$\tau$, independently of the initial number of particles in the Fock
state. 
The time $\tau_m$ at which the maximal occupation of the NO is reached
depends linearly on the number of particles $N_b$ in the initial Fock
state, and is given by $\tau_m = 0.32 N_b \hbar/t$. 

The appearance of quasi-condensates at $k=\pm \pi/2a$ can be
understood on the basis of total energy conservation. 
Given the dispersion relation of hard-core bosons on a lattice, since
the initial Fock state has a flat momentum distribution function, its
total energy is $E_T = 0$. 
If all the particles were to condense into one state, it would be to the
one with an energy $\overline{\epsilon}_k = E_T/N$. 
Taking into account the dispersion relation $\epsilon_k=-2t\cos ka$,
$\overline{\epsilon}_k=0$ corresponds to $k=\pm \pi/2a$. 
Actually, since there is only quasi-condensation in the one-dimensional case,
the argument above applies only in that the occupation of
a given state is maximized.
In addition, the minimum in the density of states at these
quasi-momenta strengthens the quasi-condensation into a single momentum
state.

Since hard-core bosons can be treated exactly 
\cite{paredes04,rigol04c,rigol05a,rigol05c,jordan28}, they are
extremely well-suited for a theoretical study of nonequilibrium 
dynamics because large systems (with hundreds to thousands of bosons)
can be examined over long times.
However, the experimental investigation is hampered by the quite stringent
requirements for the realization of such systems. 
We therefore consider here the case of finite interactions, modeled by
the one-dimensional Hubbard model 
\begin{eqnarray}
\label{BoseHubbard}
H = -t \sum_i \left( b^\dagger_i b^{}_{i+1} + h.c. \right) 
+ \frac{U}{2} \sum_i n_i \left(n_i -1 \right) \; , 
\end{eqnarray}
where $ b^\dagger_i$ and $b^{}_i$ are 
bosonic creation and annihilation operators, respectively, and 
$n_i= b^\dagger_i b^{}_i$ is the density operator.
The hard-core limit corresponds to $U \rightarrow \infty$. 
The value at which the Mott insulator appears has been estimated as 
$U_c/t \sim 3.5$ in one dimension for a commensurate density $n=1$
\cite{kuehner00}. 
Hence, all the cases considered here correspond to $ U > U_c$. 

As in the hard-core case, we start with bosons in a
Mott-insulating state spread over several lattice sites and monitor
the free expansion on a lattice. 
For the time evolution of the system, it is, in principle, necessary to
treat the whole Hilbert space of the system, restricting exact
treatments to extremely small systems.
Instead of fully diagonalizing the Hamiltonian matrix,
efficient iterative eigensolvers such as the Lanczos or the
Jacobi-Davidson procedure are commonly used \cite{cullum85,noack05},
enabling one to treat somewhat larger systems.

Unfortunately, the Lanczos method is limited by the exponential growth
of the Hilbert space as a function of the number of degrees of
freedom.
Therefore, a more efficient way of representing the relevant subspaces
for the time evolution is needed. 
Recent progress in this direction was achieved by extending 
the density matrix renormalization group (DMRG) method 
\cite{white92,white93} to treat the time evolution of correlated systems 
\cite{cazalilla02,luo03,vidal04,white04,daley04,schmitteckert04},
leading to the so-called t-DMRG.

Here we apply the t-DMRG to study the expansion of soft-core bosons out 
of a Mott insulator and compare it to the hard-core case. It will be shown 
that the essential features obtained with hard-core bosons are preserved, 
while new control possibilities are opened by tuning the strength of the 
interaction $U$. We will also show that further control of the momentum
of the emerging quasi-condensates can be achieved by introducing
a superlattice, which is obtained by superimposing an extra periodic
potential onto an already existing lattice potential. Such systems
have been recently  
realized experimentally with ultracold gases trapped on optical lattices 
\cite{peil03,strabley06}, and have been studied theoretically by
various mean field approaches \cite{buonsante04}, quantum-Monte Carlo
simulations \cite{rousseau06}, and exact diagonalization \cite{rey06}.

In Sec.\ \ref{sec:time-evo}, we discuss the theoretical treatment of
the time evolution of a 
general quantum system both within the framework of the Lanczos method
and of the t-DMRG.
The results are shown in Sec.\ \ref{sec:free-exp}, where the evolution
for parameters in the range
$6 \leq U/t \leq 40$ 
are considered. As in the hard-core case, the momentum distribution function
$n_k$ displays maxima at finite wavevectors. However, those
wavevectors are displaced to lower values of $k$ as the strength of the 
interaction is reduced. 
In Sec.\ \ref{sec:superlattice}, we study how the introduction of a
superlattice allows  
further control of the emerging quasi-condensates. For these systems
we restrict our analysis to the hard-core regime.
Finally, a concluding discussion is given in Sec.\ \ref{sec:conclusion}. 

\section{Time evolution of many-body quantum systems}
\label{sec:time-evo}
The general solution of the Schr\"odinger equation for a many-body system
can only be given in a formal way, that for a
time-independent Hamiltonian is 
\begin{equation}
|\, \psi (\tau) \, \rangle = {\rm e}^{-i H \tau} \, |\, \psi (\tau=0) \, \rangle
\; ,
\end{equation}
where $H$ is the Hamiltonian determining the evolution over a time $\tau$
from some initial time $\tau=0$. For sufficiently small systems, full 
diagonalization of $H$ is possible, and a knowledge of all the eigenvalues
and eigenstates allows for an exact determination of the evolved state.
However, for a general many-body system, the size of the Hilbert space 
grows exponentially with the number of degrees of freedom, restricting 
the system size essentially to tens of atoms. More efficient ways of 
determining the time evolution are discussed in the following subsections.

\subsection{Lanczos method}
\label{sec:Lanczos}
Here we focus on the Lanczos procedure, which can be generalized in a
straightforward way so that the time evolution of the system can be
computed without calculating all eigenstates.

The basic idea is to expand the time-evolution operator,
\begin{eqnarray}
|\, \psi (\tau + \Delta \tau) \, \rangle & = &  
{\rm e}^{-iH \Delta \tau} \, |\, \psi (\tau) \, \rangle
\simeq \sum_{n=0}^m \frac{\left(-i \Delta \tau \right)^n}{n!}
H^n  \,|\, \psi (\tau) \, \rangle \; ,
\end{eqnarray}
and to focus on the set of states
$\left\{|\, \psi (\tau) \, \rangle, H  \, |\, \psi (\tau) \, \rangle, \dots,
H^m  \, |\, \psi (\tau) \,\rangle \right\}$, 
which spans the so-called Krylov subspace.

The key idea of the Lanczos method is to obtain a basis by
orthogonalizing the vectors of the Krylov subspace only with respect to
the previous two elements of the set, leading to the recursion
relation 
\begin{eqnarray}
|\, v_{i+1} \,\rangle &=& H \, |\, v_i \,\rangle - \alpha_i \, |\, v_i \,\rangle
  -\beta_i^2 \, |\, v_{i-1} \,\rangle \label{eq:lanczos}\\
{\rm with} \quad& & \alpha_i = \frac{\langle \, v_i \, |\, \hat{H} \, |\, v_i
  \,\rangle}{\langle \, v_i \, |\, v_i \,\rangle} \, , \qquad
\beta_i^2 = \frac{\langle \, v_i \, |\, v_i
\,\rangle}{\langle \,  v_{i-1} \, |\, v_{i-1} \,\rangle}
\end{eqnarray}
for the vectors $|\, v_i \,\rangle$ of the Lanczos basis.
The projection of the Hamiltonian onto this basis set leads to a
tridiagonal matrix $T_m = V^T_m H V_m$, where $V_m$ is a rectangular
matrix containing the Lanczos vectors as column vectors.
In this way, the Hamiltonian $T_m$ can be efficiently diagonalized.

Using this approach, an approximation for the time evolution of a given
state $|\, \psi(\tau) \,\rangle$ at time $\tau$ over a small time
interval $\Delta \tau$ can be given.
For a time-independent Hamiltonian it reads
\begin{eqnarray}
|\, \psi (\tau + \Delta \tau) \,\rangle_{\rm approx} =
V^{}_m {\rm e}^{-i T_m \Delta \tau} V^T_m \, |\, \psi (\tau) \,\rangle \;
. \label{eq:lanczos_evolution} 
\end{eqnarray}
Remarkably, an exact bound can be given for the error in the
approximation 
\cite{hochbruck97}:
\begin{equation}
\mid \mid \, |\, \psi (\tau + \Delta \tau) \,\rangle -
\mid \psi (\tau + \Delta \tau) \,\rangle_{\rm approx} \mid \mid
\ \leq 12 \ \exp \left[ - \frac{\left(\rho \Delta \tau \right)^2}{16 m}\right]
\left(\frac{e \rho \Delta \tau}{4 m} \right)^m,
\end{equation}
valid for $m \geq \rho \Delta \tau/2$,  
where $\rho$ is the width of the spectrum of $H$ and $m$ is the
dimension of the Krylov subspace. 
For $\rho \Delta \tau \ll 1$ and $m$ sufficiently large (of the order
of 10), the formula above shows almost exponential convergence. 

\subsection{Adaptive time evolution with the DMRG\label{DMRG}}
The basic idea of the density-matrix renormalization group method is
to represent one or more pure states of a finite system approximately
by dividing the system in two and retaining only the $m$ most highly
weighted eigenstates of the reduced density matrix of the partial
system. 
In combination with the numerical renormalization group approach (NRG)
developed by Wilson \cite{wilson75} and the superblock algorithms
developed by White and Noack \cite{white92a}, this leads 
to a very powerful and efficient tool for the investigation of
one-dimensional strongly correlated quantum systems on a lattice.   
Here we only give a rough sketch of the method and
refer to recent reviews \cite{noack05,peschel99,schollwoeck05} for a
detailed description.

\begin{figure}
\begin{center}
\includegraphics[width=0.85\textwidth]{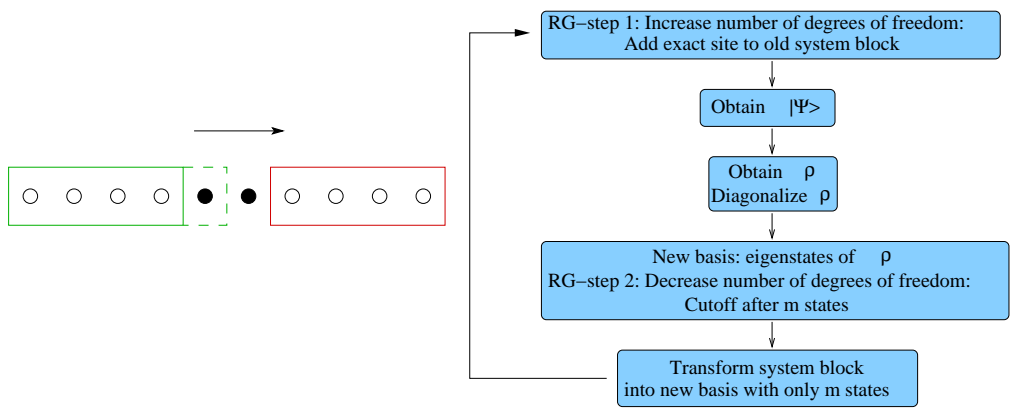}
\caption{Sketch of the lattice and flowchart of the DMRG iteration
  scheme. The lattice is shown in the usual ``superblock''
  configuration, where the left part of the lattice is the subsystem
  which is used to compute the basis of density matrix
  eigenstates. At the `dividing' bond, two ``exact'' sites are added;
  the ``sweep'' proceeds from left to right. The flowchart at the
  right shows the relevant steps of the DMRG procedure as described in
  the text.} 
\label{fig:dmrg_flowchart} 
\end{center}
\end{figure}

As depicted in Fig.\ \ref{fig:dmrg_flowchart}, the key steps are to
{\it increase} the number of degrees of freedom of the partial system
by adding sites, then to {\it decrease} the number of degrees freedom
by retaining states below a cutoff.
In this way, the method carries out a renormalization group procedure
closely related to Wilson's NRG.  

In the first step of the algorithm, a site is
added to one of the subsystems, with its Hamiltonian {\it exactly}
represented.
The Hamiltonian of the subsystem is usually represented in an {\it
  efficient} reduced basis built up from the $m$ most important
eigenstates of its reduced density matrix. 
Note that the basis is incomplete due to the truncation.
A measure of the error $\varepsilon$ introduced by the cutoff is the 
{\it discarded weight}, which measures the total weight of the
discarded states
\begin{equation}
\varepsilon = 1 - \sum\limits_{j=1}^m \lambda_j,
\end{equation}
where $\lambda_j$ is the $j^{\rm th}$ eigenvalue of the reduced
density matrix. 
After convergence is reached, one typically obtains values 
$\varepsilon < 10^{-6}$ with $m < 1000$. 
Although one is only retaining a tiny fraction of the total
Hilbert space of the system (which already for small systems can reach
a dimension of several million), the desired observables can thus be
obtained with high accuracy. 

In the second step of the iteration, the states one is interested in
are obtained. 
These states are called ``target states''.
In the original ground-state algorithm, these are the ground state and
the few lowest lying excited states of the system, which are obtained
by diagonalizing the Hamiltonian of the {\it total} system, e.g., by
carrying out the Lanczos diagonalization algorithm described in the
introduction.
However, states other than the eigenstates of the Hamiltonian may be
obtained in this step.
This flexibility is crucial for the time-evolution algorithms, in which
the time-evolved state is obtained by other means than
solving an eigenvalue problem. 

In the third step, the new effective basis is obtained by diagonalizing
the reduced density matrix of the extended subsystem, given by
\begin{equation}
\rho_{\rm subsystem} = {\rm Tr}_{\rm rest} \left(\sum\limits_i n_i \, |\, \psi_i
\,\rangle \langle\, \psi_i \, | \, \right) \ , \quad \sum n_i = 1 \; ,
\end{equation}
where the sum goes over all target states.
In step four, only the $m$ eigenstates with the largest eigenvalues
are kept. 
The operators needed to represent the Hamiltonian of the subsystem, to form
the pieces of the Hamiltonian connecting subsystems, and to calculate
observables are transformed into this new reduced basis. 
This effective Hamiltonian of the subsystem is now the starting point 
for step one of the next iteration. 
In this way, every step improves the accuracy of the
obtained eigenstates and energies by improving the reduced basis used
for the representation of the target states. 

DMRG iteration schemes are usually divided into two classes. 
In the so-called {\it infinite-system} algorithm,
the system grows at each step.
This can be used to build up the system up to a desired lattice size.
In the {\it finite-system} algorithm,
the size of the lattice is fixed, and the ``dividing bond'', i.e., the
position at which the system is cut in two parts, is moved
from the right end of the lattice to the left end and back (other
variations are possible). 
This is called a ``sweep''.
In order to obtain the ground state (and optionally the lowest lying
excited states) with a high accuracy, the sweeps are iterated until
convergence is reached.
The calculation can be significantly sped up if the diagonalization
in step 2 of the DMRG procedure is started with a good initial guess
for the wave function.
Such an initial guess can be constructed using the
so-called ``wave function transformation'', which approximately
transforms the wave function obtained from the previous finite-system
step into the basis of the current superblock configuration.
As we will see next, the wave function transformation
also plays a key role in the adaptive t-DMRG schemes.

The main difficulty in calculating the time evolution using the DMRG
is that the restricted basis determined at the beginning of the
time evolution is not able, in general, to represent the state well at
later times \cite{luo03} because it covers a subspace of the
total Hilbert space which is not appropriate to properly
represent the state at the next time step. 
Since both the Hamiltonian and the wave function $|\, \psi(\tau)\,
\rangle$ at time $\tau$ are represented in an incomplete basis, the
result for the next time step $|\, \psi(\tau+\Delta\tau)\,\rangle$ will 
have additional errors because the reduced basis is not an optimum
representation for this state. 
In order to minimize these errors, it is necessary to form 
a density matrix whose $m$ most important eigenvectors are
``optimal'' for the representation of the state $|\, \psi(\tau)
\,\rangle$, as well as for $|\, \psi(\tau+\Delta \tau)\,\rangle$ in the
reduced Hilbert space.  
The most straightforward approach is to mix {\it all} time steps
$|\, \psi(\tau_i)\,\rangle$ into the density matrix
\cite{luo03,schmitteckert04}. 
However, this can be extremely costly computationally.
A more efficient way is to {\it adapt} the density matrix at each time
step.

\begin{figure}
\includegraphics[width=0.5\textwidth]{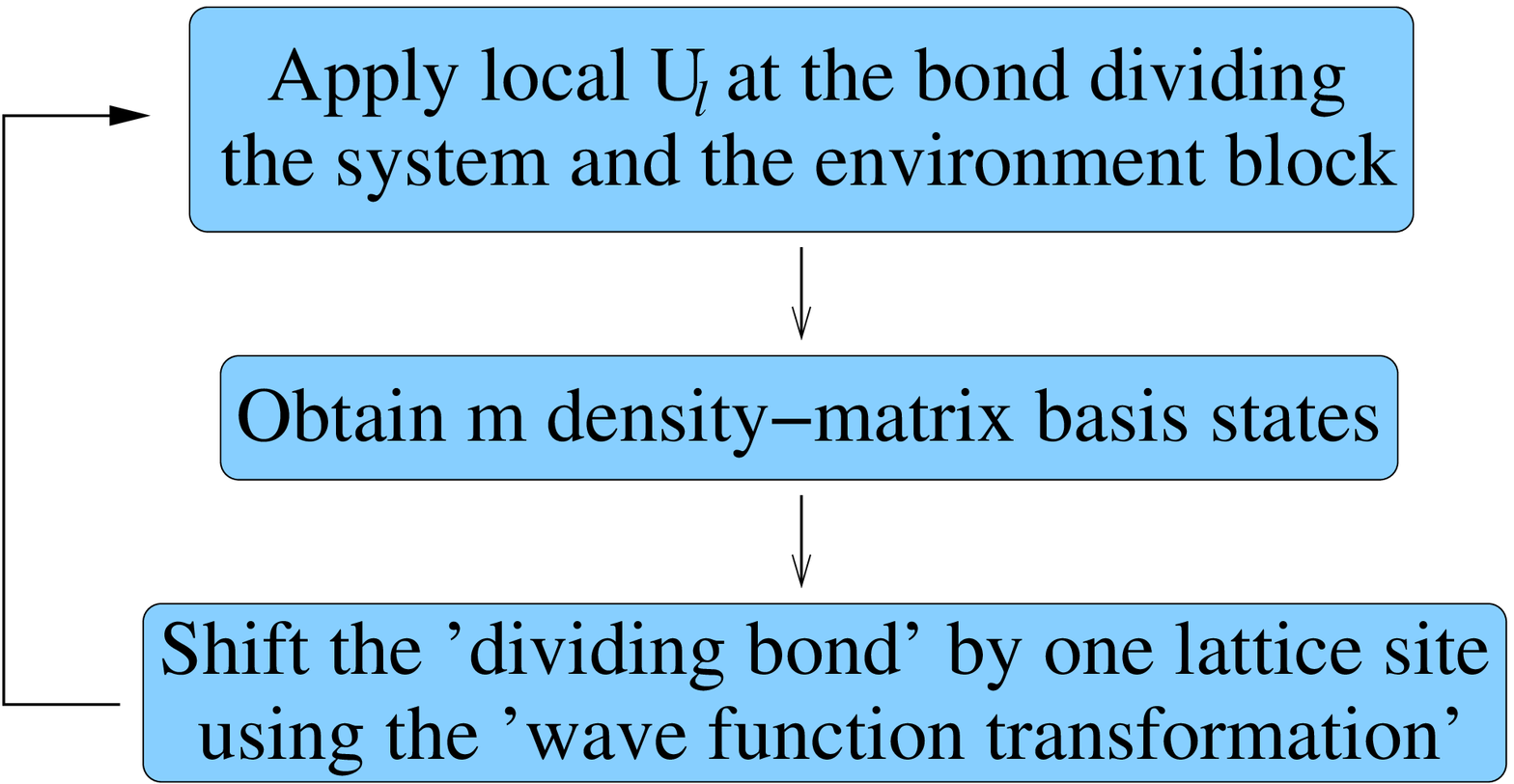}
\includegraphics[width=0.5\textwidth]{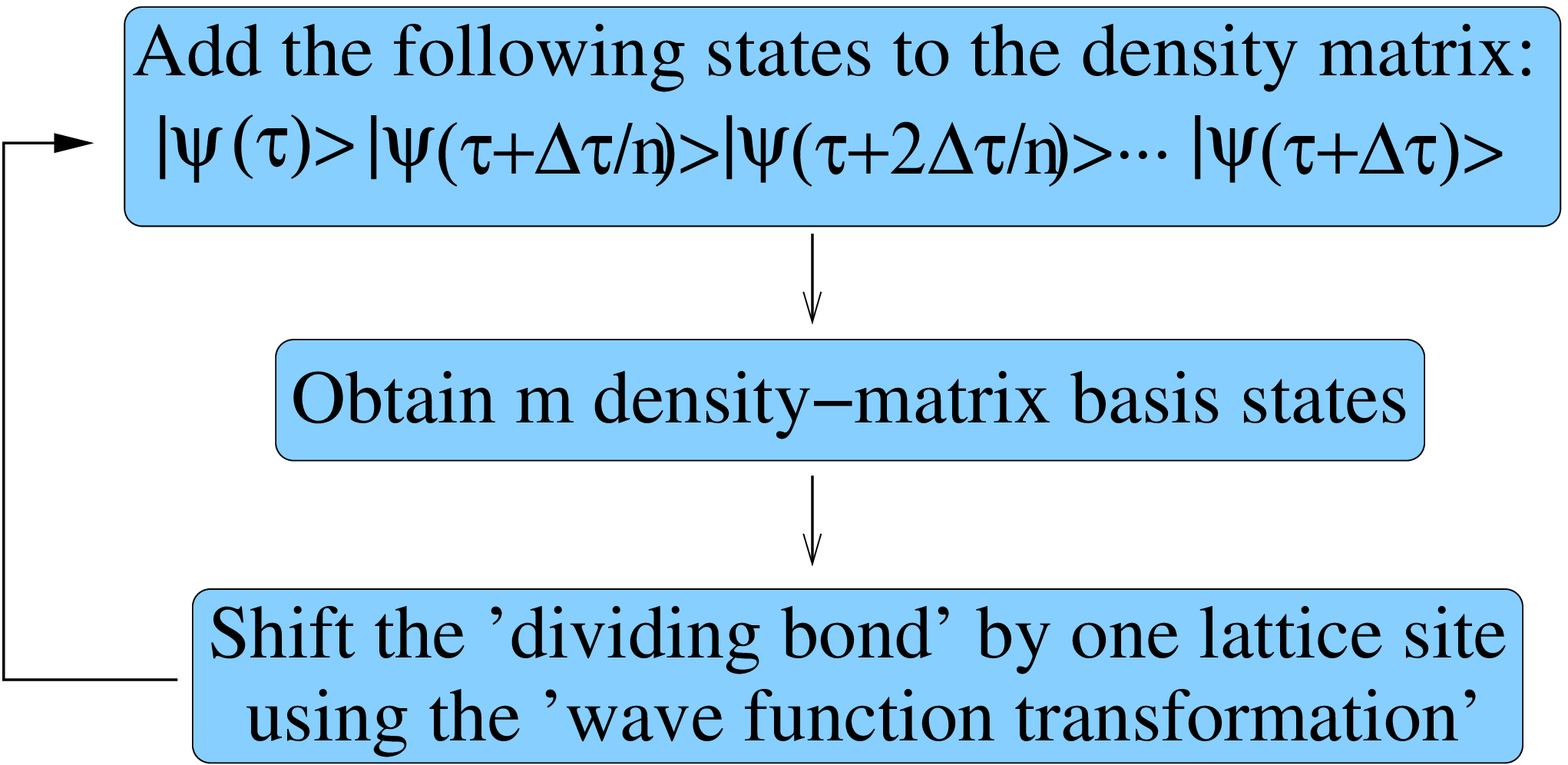}
\caption{The flowcharts of the t-DMRG schemes described in the
  text. On the left, the flowchart for the Trotter variant is
  sketched. On the right, the scheme used for the
  Lanczos variant is shown.}
\label{fig:tdmrg}
\end{figure}

An approach for adaptive time evolution based on the Trotter-Suzuki
\cite{suzuki76} decomposition of the time-evolution operator was
developed in Refs. \cite{vidal04,white04,daley04}.
The idea is to split up the time-evolution operator in {\it local}
time-evolution operators $U_l$ acting only on the {\it bond} $l$. 
For lattice Hamiltonians containing only terms connecting
nearest-neighbor sites, this is easily obtained using the
Trotter-Suzuki decomposition, which in second order is given by 
\begin{equation}
  e^{-i \Delta \tau H} \approx e^{-i \Delta \tau H_{\rm even}/2} \,
                               e^{-i \Delta \tau H_{\rm odd}}    \,
                               e^{-i \Delta \tau H_{\rm even}/2} \, .
\label{eq:trotter}
\end{equation}
Here $H_{\rm even}$ and $H_{\rm odd}$ is the part of the Hamiltonian
containing terms on even and odd bonds, respectively.
Since each bond term $H_l$ within $H_{\rm even}$ or $H_{\rm odd}$
commutes, $e^{-i \Delta \tau H}$ can then be factorized into terms
acting on individual bonds.
As depicted in Fig.\ \ref{fig:dmrg_flowchart}, in the DMRG procedure
usually two sites are treated exactly, i.e., the entire Hilbert space
of the two sites is included.
The Trotter variant of the t-DMRG exploits this feature by applying
$U_l = e^{-i \Delta \tau H_{l}}$
at the bond given by the two ``exact'' sites.
In this way, the time-evolution operator has no further
approximations other than the error introduced by the Trotter
decomposition.
In particular, the error introduced by the cutoff is avoided. 
The wave function of the lattice is then updated by performing one
complete sweep over the lattice and applying $U_l$ at the ``dividing
bond''.
In this way, only one wave function must be retained and it is
possible to work with the density matrix for a pure state.
The flowchart is sketched in Fig.\ \ref{fig:tdmrg}.

However, the method is restricted to systems with local or
nearest-neighbor terms in the Hamiltonian. 
A more general basis adaption scheme aims at adapting the density
matrix basis by approximating the density matrix for a time interval 
\cite{feiguin05},  
\[
\rho_{\Delta\tau} = \int\limits_{\tau}^{\tau+\Delta\tau}
|\,\psi(\tau')\,\rangle \,\langle\, \psi(\tau')\,| \; d\tau'.
\]
The integral is approximated by adding a few {\it intermediate}
time steps within the time interval $[\tau, \tau+\Delta\tau]$.
In Ref. \cite{feiguin05}, the intermediate time steps are obtained
using a  Runge-Kutta integration scheme and using 4 or 10 intermediate
time steps. 
Here we instead obtain the intermediate
time steps using the Lanczos method described in Sec.\ \ref{sec:Lanczos}. 
This can be done easily because the Hamiltonian of the system is
usually constructed anyway in the DMRG scheme. 
Within the restricted basis, the Lanczos iteration, 
Eq.\ (\ref{eq:lanczos}),  can then be performed, leading to the desired
intermediate time steps computed using Eq.\ (\ref{eq:lanczos_evolution}). 
Using this approach, we find that, if the time step is small enough, it
is sufficient to retain only the target state $|\, \psi(\tau+\Delta\tau)
\,\rangle$, so that one can work with a pure-state density matrix like
in the Trotter approach.
For larger time steps, it is important to mix at least the states
$|\, \psi(\tau)\,\rangle$ and $|\, \psi(\tau+\Delta\tau) \,\rangle$ into the
density matrix.
We find that it is sufficient to perform only one half-sweep in order
to adapt the restricted basis, which is the minimum requirement for
updating the basis on the complete lattice. 
The flowchart of this approach is sketched in Fig.\ \ref{fig:tdmrg}. 
With this approach, it is, in principle, possible to treat more
general Hamiltonians, as long as they can be treated accurately using
the DMRG.
However, due to the fact that, in general, one cannot work with a pure
state density matrix, the dimension $m$ of the restricted basis needed
to obtain a certain discarded weight during the time
evolution is larger compared to the Trotter approach described above,
making this variant slower.  The errors in both adaptive schemes
for intermediate and long times are comparable.
For the problem at hand, we therefore choose the Trotter variant (in 
second order) of the adaptive time-dependent DMRG.

In this work, we control the error during the time evolution by
fixing the discarded weight and varying the number of basis states
kept. 
By keeping a maximum of $m=800$ density matrix eigenstates, we obtain
discarded weights smaller than $5\cdot10^{-7}$ during the time
evolution. 

\begin{figure}
\begin{center}
\includegraphics[width=0.5\textwidth]{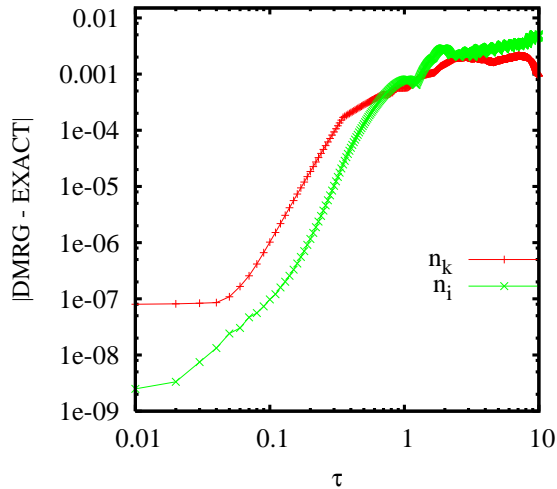}
\caption{Error analysis of the t-DMRG obtained by comparing the
  results for the  
  density $\langle n_i \,\rangle$ and of the momentum distribution
  $n_k$ of an initial Fock state with 10 hard-core
  bosons on a lattice with 50 sites with exact results obtained
  using the Jordan-Wigner transformation. The t-DMRG results were
  calculated using the Trotter variant (second order) of the method, 
keeping up to
  $m=200$ basis states.}
\label{fig:tDMRG_hcb_error}
\end{center}
\end{figure}

The initial state has zero density over a wide region in the system. 
This may lead to difficulties for the DMRG.
In order to control this, we compare the time evolution of an initial
Fock state of hard-core bosons obtained using the t-DMRG with the
exact results from the Jordan-Wigner transformation.
The maximum error as a function of time is plotted in
Fig. \ref{fig:tDMRG_hcb_error}. 
As shown in this figure, the maximum deviation from the exact results
is smaller than 0.01 for all times considered. 

\section{Free expansion of soft-core bosons from a Mott insulator}
\label{sec:free-exp}
We consider here the free expansion of $N_b$ interacting bosons described by 
the Hamiltonian (\ref{BoseHubbard}) on a one-dimensional
lattice with $L$ sites, lattice constant $a$ and open boundary conditions. 
In the cases considered
here, we take $N_b = 20$ and $L=60$. Due to memory limitations, it is not 
possible to allow for all possible occupations of a given site. In general,
the maximal number of bosons per site needed to have an accurate description 
of the system increases as $U$ decreases. In all the cases treated here a 
maximum of three bosons per site was sufficient. Even at the smallest
interaction studied here ($U/t = 6$), no appreciable difference was observed
when the cutoff was changed from 
3 to 4 bosons per site. Since the system
becomes more dilute in the course of the free expansion, the
limitation in the number of bosons per site becomes even less important
at later times. The time sequences shown are all limited to times 
shorter than the time it takes the matter wave to reach the boundary of the 
system. 

\begin{figure}[h]
\begin{center}
\includegraphics[width=\textwidth]{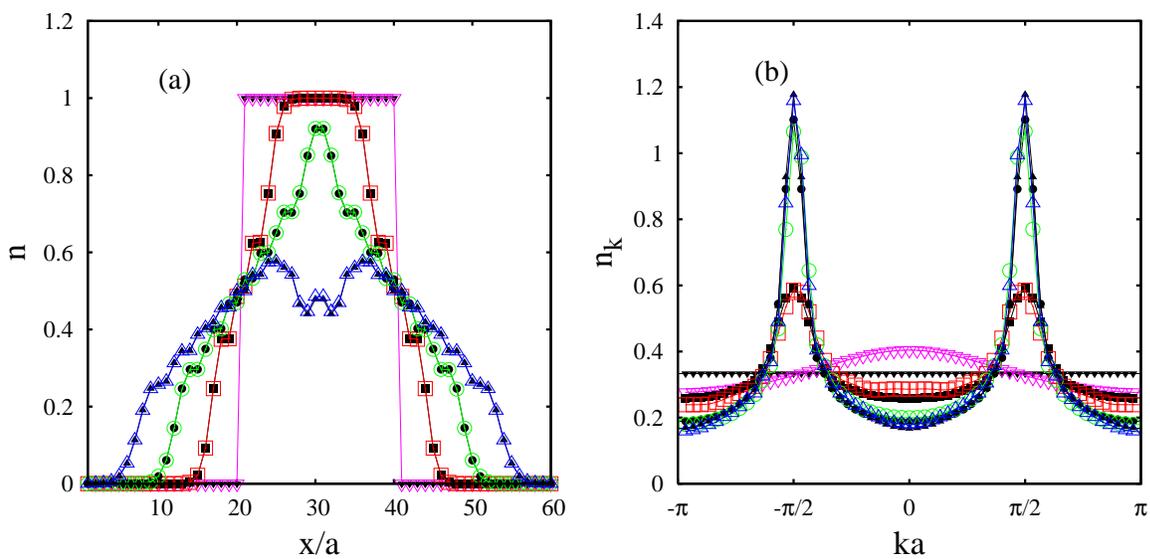}
\end{center}
\caption{Comparison between hard-core bosons (filled black symbols) and 
soft-core bosons (unfilled colored symbols) with $U/t=40$ for (a)
density and (b) $n_k$ at time $\tau=0$ (\textcolor{magenta}{$\nabla$}), 
$\tau=2.52$ (\textcolor{red}{$\square$}), 
$\tau=4.98$ (\textcolor{green}{$\bigcirc$}),
and $\tau=7.5$ (\textcolor{blue}{$\triangle$}) in units of $1/t$.
}
\label{U40}
\end{figure}
Figure \ref{U40} shows a comparison of the density (Fig.\ \ref{U40}(a)) and
the momentum distribution function (Fig.\ \ref{U40}(b)) for a system 
with $U/t=40$ with hard-core bosons at four different times. 
At such high values of the interaction, it is expected that the particles
behave as hard-core bosons. In fact, there is no noticeable difference in the 
density profiles at any time. However, the momentum distribution functions at 
$\tau=0$ show clear differences. While hard-core bosons are equally 
distributed over all momenta, the soft-core case has a maximum at
$k=0$, showing that even in a Mott insulator, the fluctuations of the number
of particles at each side populate that state preferentially. Nevertheless, 
after the Mott insulator is allowed to expand, the difference between both 
systems becomes barely noticeable. Already at the second time shown in 
Fig.\ \ref{U40}, where a Mott plateau still exists at the center of the cloud
(Fig.\ \ref{U40}(a)), the momentum distribution functions of the hard- and 
soft-core bosons are very close to each other. This is expected because
the constraint of a hard-core should become less
relevant when the system is diluted. 

\begin{figure}
\begin{center}
\includegraphics[width=\textwidth]{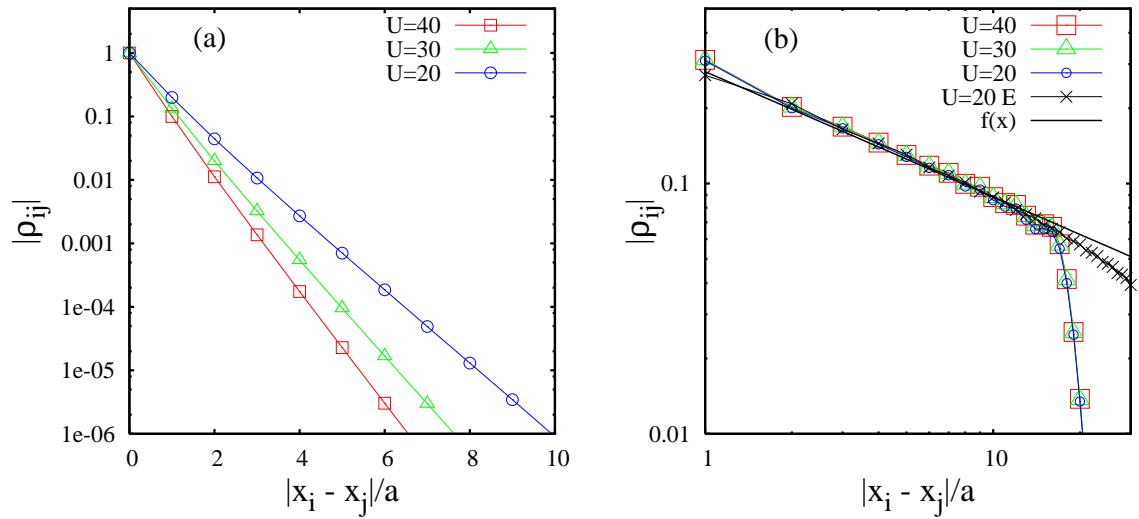}
\caption{
(a) Modulus of the one-particle density matrix $\rho_{ij}$ {\it versus} 
distance $\mid x_i - x_j \mid$ in units of the lattice constant $a$ at
$\tau=0$. (b) The same quantity at time $\tau=7.38$, where the peaks
in $n_k$ are fully developed. Crosses (U=20 E) correspond to the 
one-particle density matrix in equilibrium for the same number of particles 
and system size. The black line corresponds to 
$f(x) \sim 1/\sqrt{\mid x_i - x_j\mid}$.
}
\label{Decay}
\end{center}
\end{figure}
In order to see the establishment of coherence explicitly, we
examine the spatial behavior of the one-particle density matrix.
We would expect a change from an exponential decay in the
Mott-insulating state to a power law behavior if (quasi-)co\-herence
emerges.
Figure \ref{Decay} shows the spatial behavior of the one-particle
density matrix both at time $\tau=0$ (Fig.\ \ref{Decay}(a)), when bosons 
are in a Mott-insulating state, and at time $\tau \sim 7.5$ 
(Fig.\ \ref{Decay}(b)), 
when the peaks around $k=\pm \pi/2a$ are well established. 
Figure \ref{Decay}(a) shows the spatial dependence  of the one-particle 
density matrix measured from the center of the bosonic region in a 
semi-logarithmic plot for different values of $U$. As expected for a 
Mott insulator, an exponential decay with a correlation 
length that shortens as $U$ is increased is observed. Figure
\ref{Decay}(b) shows the decay of the one-particle density matrix on a
log-log scale at a time
long enough so that the peaks around $k=\pm \pi/2a$ are fully developed.
\begin{figure}
\begin{center}
\includegraphics[width=\textwidth]{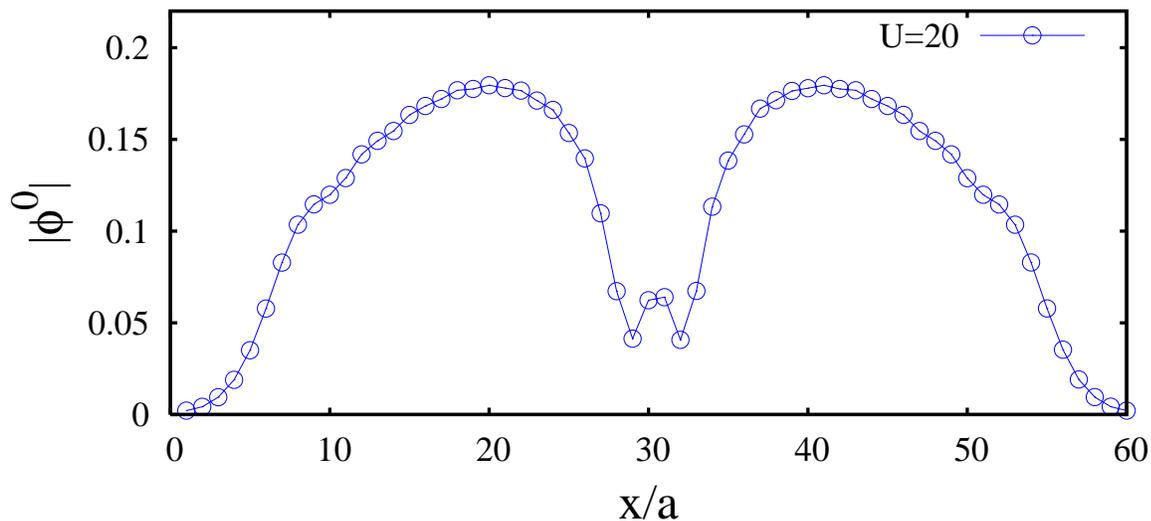}
\caption{Spatial dependence of the lowest NO at time $\tau=7.38$ for 
$U/t=20$.  
}
\label{NaturalOrb}
\end{center}
\end{figure}
The evaluation was made in the part of the system where the lowest NO is 
appreciable, i.e.\ in the region of the system where a well developed 
quasi-condensate can be expected.
Figure \ref{NaturalOrb} shows the spatial dependence of the lowest NO
at the same time as in Fig.\ \ref{Decay}(b). The correlations in 
Fig.\ \ref{Decay}(b)  were measured 
from the site $x_j=37 a$ (a position where the NO is well developed)
and  with $x_i > x_j$.  
For comparison, we superimposed the one-particle density matrix for 
$U/t = 20 $, $N_b = 20 $, and $L=60$ in equilibrium, where, due to the
lower density with respect to the initial state in Fig.\ \ref{Decay}(a), 
a quasi-condensate exists. Over the distances where 
the NO has an appreciable value, no difference with the corresponding quantity
in equilibrium can be noticed.
It can be clearly seen that the one-particle density matrix has developed
a power-law decay (the same as the one in the system in equilibrium)
at the later time, with a power that approaches the
one of hard-core bosons. 
Unfortunately, the
expansion of the cloud and the total system size are not as large in
the soft-core (as seen in the departures from the power-law due to 
finite-size effects) as in the
hard-core case, so that the exponent of the power-law decay cannot be as 
accurately determined. Nevertheless, it is clear 
that a change from an exponential to a power-law decay takes place, indicating
that a quasi-coherent matter wave has developed.

\begin{figure}
\begin{center}
\includegraphics[width=\textwidth]{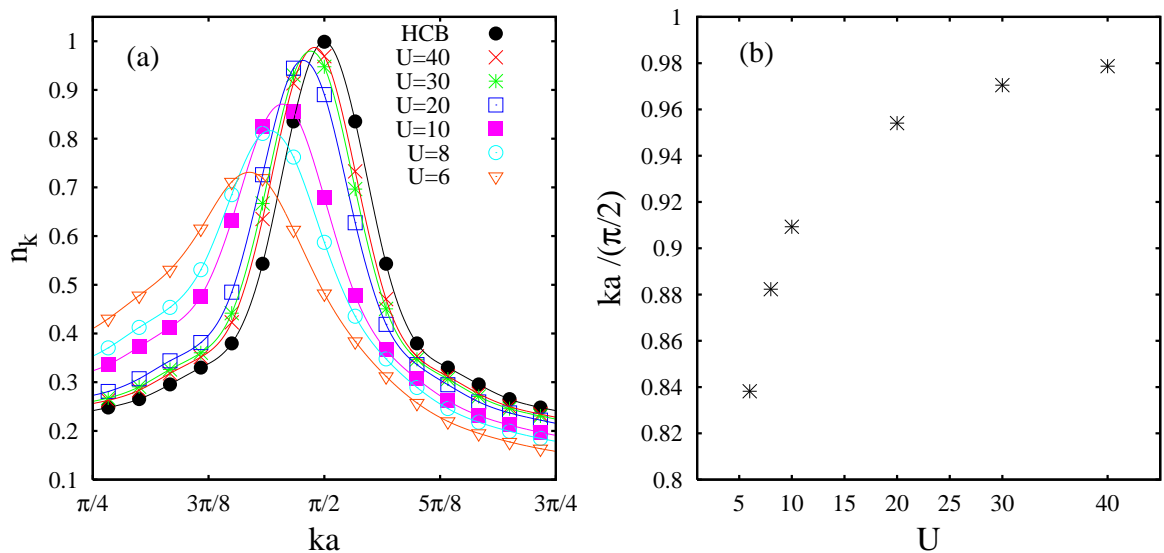}
\caption{
(a) The momentum distribution $n_k$ at time $\tau=4.5/t$ for
  different values of $U/t$. The symbols  
correspond to t-DMRG results on a lattice with $L=60$, while the lines
in the respective colors are spline interpolations.
(b) The position of the peaks of the interpolated splines in $n_k$ as
a function of $U/t$. 
}
\label{Smooth}
\end{center}
\end{figure}

Finally, we discuss the behavior of the expansion for smaller values of the
interaction $U$. Although at $U/t=40$ the momentum distribution function 
closely follows the shape of $n_k$ of hard-core bosons, a tiny
asymmetry can be seen in Fig.\ \ref{U40}(b) around the peaks at $k=\pi/2a$.  
Such an asymmetry indicates that the maximum of $n_k$ is not exactly at 
$k=\pi/2a$, but is shifted slightly. A more detailed analysis for 
$6 \leq U/t \leq 40$ is presented in Fig.\ \ref{Smooth}.
Figure \ref{Smooth}(a) shows $n_k$ around $k=\pi/2a$ with the data points 
from t-DMRG denoted by symbols with spline interpolations between them. 
It is clearly seen that the maximum of $n_k$ is displaced to smaller
momenta as $U$ decreases. The spline interpolation allows for a better
determination 
of the maxima in $n_k$, since a denser set of $k$-points 
corresponds to having a much longer lattice in a physical realization. 
On the other hand, the actual set of $k$-points in the t-DMRG simulation 
corresponding
to $L=60$ is dense enough to allow for a smooth interpolation without
introducing artefacts due to the spline procedure. Figure \ref{Smooth}(b)
shows the location of the maxima of $n_k$ as a function of U in units of
$2a/\pi$, giving a guide for a fine tuning of the wavelength of the matter 
wave via the interaction strength.

The results in this section show that the main feature found for the
case of hard-core bosons \cite{rigol04c,rigol05c}, namely, the
emergence of a quasi-coherent matter wave from a Mott insulator, persists
under more general conditions.
The wavelength of the matter wave is determined primarily by the
underlying lattice, on which the expansion takes place, as in the case
of hard-core bosons.
Additionally, finer tuning is 
possible by regulating the interaction strength of the bosons.
This means that on an optical lattice the wavelength of the 
corresponding laser beam would essentially determine the momentum of the 
matter wave and a fine tuning can be reached by varying its intensity.
The next section discusses further control possibilities 
by introducing more complex structures in the optical lattice.

\section{Expansion in a superlattice}
\label{sec:superlattice}
So far we have studied how finite values of the on-site interactions
between bosons modify the behavior already known in the hard-core limit. 
The general finding has been that the physics is similar, and that an 
emergence of quasi-condensates can be obtained experimentally 
for a wide range of finite repulsive interactions. In this section,
we analyze how the introduction of a superlattice potential allows a
further  
degree of control over the system and can lead to a richer momentum
distribution of the expanding cloud of bosons. We will restrict the
analysis to the  
hard-core limit keeping in mind its relevance to the soft-core regime.

In the presence of a superlattice 
potential, the hard-core boson Hamiltonian becomes
\begin{equation}
\label{HamHCB} H = -t \sum_{i} \left( b^\dagger_{i} b^{}_{i+1}
+ \mathrm{H.c.} \right) + A \sum_{i} \cos {2 \pi i \over \ell} n_i,
\end{equation}
with the additional on-site constraints
\begin{equation}
\label{ConstHCB} b^{\dagger 2}_{i}= b^2_{i}=0, \quad  
\left\lbrace  b^{}_{i},b^{\dagger}_{i}\right\rbrace =1, 
\end{equation}
which exclude double or higher occupancy. The bosonic 
creation and annihilation operators at site $i$ 
are denoted by $b^{\dagger}_{i}$ and $b^{}_{i}$, respectively, 
and the local density operator by $n_i=b^{\dagger}_{i}b^{}_{i}$.
The brackets in Eq.\ (\ref{ConstHCB}) apply only to on-site 
anticommutation relations; for $i\neq j$, these operators 
commute as usual for bosons; $[b^{}_{i},b^{\dagger}_{j}]=0$. 
In Eq.\ (\ref{HamHCB}), the hopping parameter is denoted by $t$
and the last term represents the superlattice potential with
strength $A$ and  $\ell$ sites per unit cell.

Using the Jordan-Wigner transformation \cite{jordan28}
\begin{equation}
\label{JordanWigner} b^{\dag}_i=f^{\dag}_i
\prod^{i-1}_{\beta=1}e^{-i\pi f^{\dag}_{\beta}f^{}_{\beta}},\quad
b_i=\prod^{i-1}_{\beta=1} e^{i\pi f^{\dag}_{\beta}f^{}_{\beta}}
f_i \ ,
\end{equation}
one can map the HCB Hamiltonian onto that of 
noninteracting spinless fermions,
\begin{eqnarray}
\label{HamFerm} H_F =-t \sum_{i} \left( f^\dagger_{i}
f^{}_{i+1} + \mathrm{H.c.} \right)+ A \sum_{i} \cos 
{2 \pi i \over \ell} n^f_i\, ,
\end{eqnarray}
where $f^\dagger_{i}$ and $f_{i}$ are the creation and
annihilation operators for spinless fermions at site 
$i$, and $n^f_{i}=f^\dagger_{i}f^{}_{i}$ is the local 
particle number operator. For periodic systems with $N$ 
lattice sites, one needs to consider that
\begin{equation}
b^\dagger_{1} b^{}_{N}=-f^\dagger_{1}f^{}_{N}\ 
\exp\left( i\pi \sum^N_{\beta=1}n^f_{\beta}\right) ,
\end{equation}  
so that when the number of particles in the system 
[$\sum_{i}\langle n_i \rangle=\sum_{i}\langle n^f_i\rangle=N_b$]
is odd, the equivalent fermionic Hamiltonian satisfies periodic
boundary conditions; whereas, if $N_b$ is even, antiperiodic 
boundary conditions are required.

The above mapping to noninteracting fermions allows one to 
realize that the presence of an additional periodic potential
opens gaps at the edges of the reduced Brillouin zones. 
This implies that new insulating phases appear at fractional 
fillings $n_i=i/\ell$, with $i=1,\ldots, \ell-1$, in addition to the
insulating phase at $n=1$ which is present in the  
absence of the superlattice.

In the following, we address the question of what happens
during the evolution of initially prepared insulating states when 
they are allowed to expand in a superlattice. For simplicity, we 
restrict our analysis to the case $\ell=2$. (The generalization to larger 
values of $\ell$ is straightforward.) To study these systems, we follow the
exact approach already described in detail in
Refs.\ \cite{rigol04c,rigol05a,rigol05c}.

For $\ell=2$, a band gap $\Delta=2A$ opens at $k=\pi/2a$. The dispersion 
relation for the two bands reads
\begin{equation}
\epsilon_{\pm}(k) = \pm \sqrt{ 4 t^2 \cos^2(ka) + A^2} \, ,
\end{equation}
where `$+$' stands for the upper band and `$-$' for the lower one. 
One can then calculate the group velocity at each 
momentum as
\begin{equation}
\nu^g_\pm=\frac{\partial \epsilon_{\pm}(k)}{\partial k}
=\mp \frac{2t^2 \sin(2ka)}{\sqrt{ 4 t^2 \cos^2(ka) + A^2}},
\label{velocity}
\end{equation}
which means that the wave vectors at which the maximum group velocity 
occurs satisfy 
\begin{equation}
\cos^2(k_ma)=\frac{\sqrt{1+\frac{4t^2}{A^2}}-1}{\frac{4t^2}{A^2}}\, .
\label{extremum}
\end{equation}
For these values of $k$, the density of states attains its minimum 
values. 
Hence, under the appropriate initial conditions,  
we should expect the quasi-condensates to emerge at these values of
$k$ rather than at $k=\pi/2$, as in the absence  
of the superlattice ($A=0$).

\begin{figure}[h]
\begin{center}
\includegraphics[width=\textwidth]{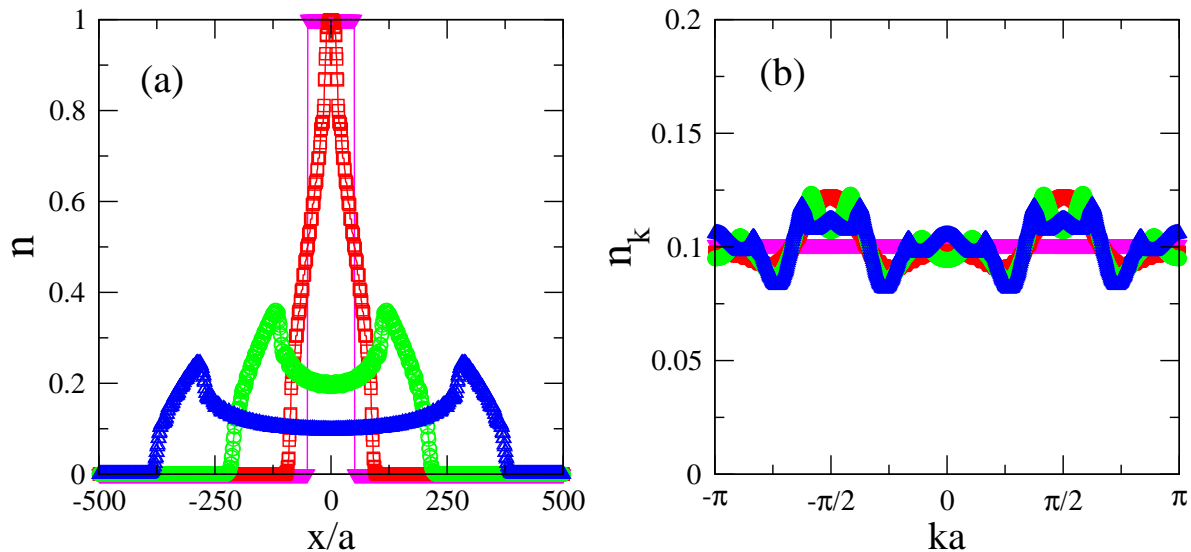}
\caption{Evolution of (a) density and (b) momentum 
profiles of 101 HCB's, initially prepared in an insulating state with 
density one, on 1000 lattice sites. The superlattice
parameters are $\ell=2$ and $A=2t$. The times are 
$\tau=0$ (\textcolor{magenta}{$\nabla$}), 
$50\hbar/t$ (\textcolor{red}{$\square$}), 
$200\hbar/t$ (\textcolor{green}{$\bigcirc$}), and 
$400\hbar/t$ (\textcolor{blue}{$\triangle$}).
}
\label{PerfilesN1.0}
\end{center}
\end{figure}

Since for $\ell=2$ the only insulating phases occur at full filling and at
half filling, we start studying the evolution of an insulating state 
initially prepared with one particle per lattice site, like in 
the previous sections and in Refs.\ \cite{rigol04c,rigol05c}. 
The expansion of such a state with 101 HCB's is shown 
in Fig.\ \ref{PerfilesN1.0}. While the evolution of the density 
[Fig.\ \ref{PerfilesN1.0}(a)] is very similar to that in the absence 
of the superlattice \cite{rigol04c,rigol05c}, the evolution of the 
momentum distribution function is completely different 
[Fig.\ \ref{PerfilesN1.0}(b)]. No peaks appear in $n_k$, in contrast 
to the case analyzed in Refs.\ \cite{rigol04c,rigol05c} and in the 
previous sections. This is because in the superlattice the mean energy 
per particle ($\overline{\epsilon}=0$) for the fully filled insulating 
state lies in the band 
gap, and the states with the closest energy are the ones with momentum 
$k=\pm \pi/2$, which have $\nu^g_\pm=0$ and the maximum density of 
states, i.e., exactly the opposite of the case without the superlattice.

A scenario in the superlattice that is closer to the one in a Mott 
insulator with $n=1$ (and finite $U$) in the absence of the superlattice, 
is the one in which the initial state is prepared with a mean density 
of 0.5. Such state has short-range (exponentially 
decaying) one-particle correlations like the ones seen in Fig.\ \ref{Decay}. 
In addition, its mean energy per particle lies within the lowest band,
where the density of states is finite. 
\begin{figure}[h]
\begin{center}
\includegraphics[width=\textwidth]{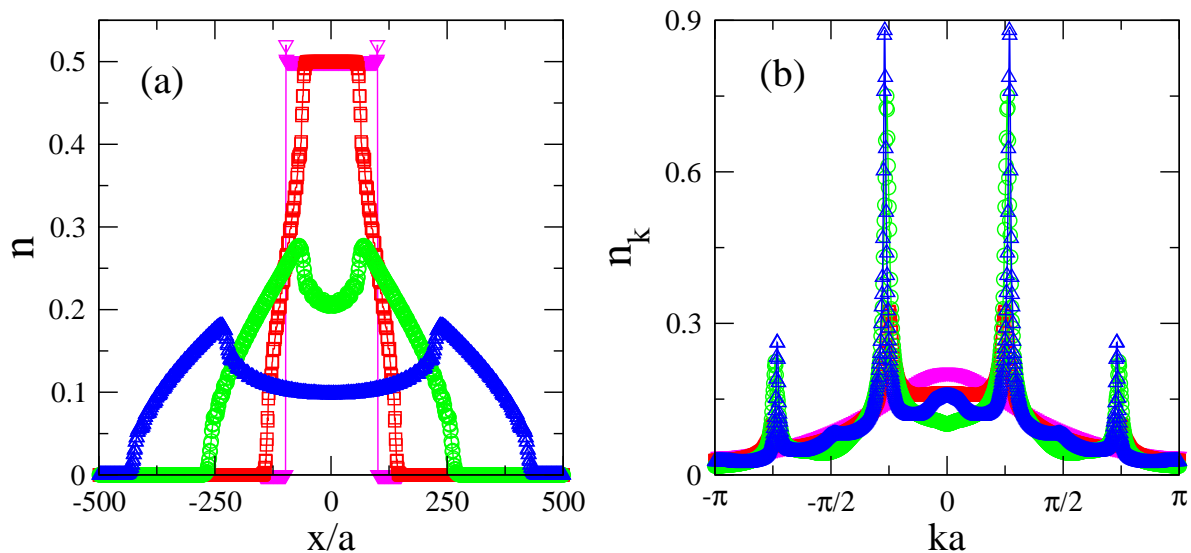}
\caption{Evolution of (a) the mean density per unit cell and (b) momentum 
profiles of 100 HCB's, initially prepared in an insulating state with
a mean density of 0.5, on 1000 lattice sites. The superlattice
parameters are $\ell=2$ and $A=2t$. The times are 
$\tau=0$ (\textcolor{magenta}{$\nabla$}), 
$50\hbar/t$ (\textcolor{red}{$\square$}), 
$200\hbar/t$ (\textcolor{green}{$\bigcirc$}), and 
$400\hbar/t$ (\textcolor{blue}{$\triangle$}).
}
\label{PerfilesN0.5}
\end{center}
\end{figure}

In Fig.\ \ref{PerfilesN0.5} we show the evolution of the mean density 
per unit cell and momentum profiles during the expansion of a state 
prepared in a half-filled box with 100 HCB's and $A=2t$. As can 
be seen in Fig.\ \ref{PerfilesN0.5}(b), the initial momentum distribution 
is not flat like the one in Fig.\ \ref{PerfilesN1.0}(b). 
Its maximum at $k=0$ signals the presence of short-range correlations 
like in the Mott insulator of Fig.\ \ref{U40}. Remarkably, during the 
expansion of this state, sharp peaks appear in $n_k$ at $ka=\pm 0.87$ and
$ka=\pm 2.27$, which are the momenta for which the group velocity is 
maximum and the density of states has a minimum, following 
Eq.\ (\ref{extremum}). 

The peaks in $n_k$ signal the emergence of quasi-condensates of 
HCB's at finite momentum. This can be seen by studying the natural 
orbitals ($\phi^\eta$), which can be considered to be effective 
single-particle states in interacting systems, and are defined as 
the eigenfunctions of the one-particle density matrix
$\rho_{ij}$ \cite{penrose56},
\begin{equation}
\label{NatOrb}
\sum^N_{j=1} \rho_{ij}(\tau)\phi^\eta_j(\tau)=
\lambda_{\eta}(\tau)\phi^\eta_i(\tau),
\end{equation}
with occupations $\lambda_{\eta}$.
In dilute higher dimensional 
gases, where only the lowest natural orbital (the highest occupied one) 
scales $\sim N_b$, the occupation of this orbital can be regarded as
the BEC order  parameter, i.e., the condensate \cite{leggett01}.

\begin{figure}[h]
\begin{center}
\includegraphics[width=\textwidth]{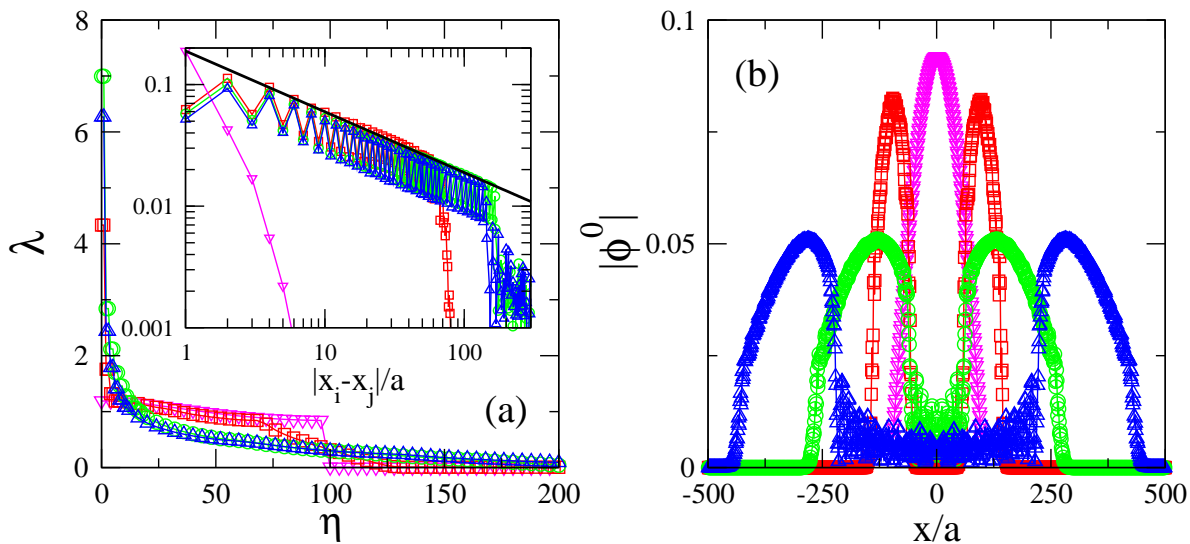}
\caption{Evolution of (a) the natural orbital occupations and 
(b) the modulus of the lowest natural orbital wave function (averaged in the 
unit cell) for 100 HCB's initially prepared in an insulating 
state with mean density of 0.5 on 1000 lattice sites. The 
superlattice parameters are $\ell=2$ and $A=2t$. The inset in (a) shows 
the initial exponential decay of one-particle correlations 
(averaged per unit cell), and its conversion to a power-law 
decay in the region where the quasi-condensate forms. 
The black line corresponds to $1/\sqrt{\mid x_i - x_j \mid}$. 
The times are $\tau=0$ (\textcolor{magenta}{$\nabla$}), 
$50\hbar/t$ (\textcolor{red}{$\square$}), 
$200\hbar/t$ (\textcolor{green}{$\bigcirc$}), and 
$400\hbar/t$ (\textcolor{blue}{$\triangle$}).
}
\label{NatOrbN0.5}
\end{center}
\end{figure}

In Fig.\ \ref{NatOrbN0.5}(a) we show the values of $\lambda_{\eta}$ for 
the 200 highest occupied orbitals after the same expansion times as in
Fig.\ \ref{PerfilesN0.5}. One can see that during the expansion 
the lowest natural orbital becomes ``highly'' populated with of the 
order of $\sqrt{N_b}$ particles. This is because quasi-long range 
correlations develop in the system, as shown in the inset of 
Fig.\ \ref{NatOrbN0.5}(a). The power-law decay of the one-particle 
correlations is $\sim 1/\sqrt{|x_i-x_j|}$, i.e., the same form that 
appears in the absence of the superlattice \cite{rigol04c,rigol05c}, 
and that has been proven to be universal in the ground state
\cite{rigol04b,rigol05b}. The wave function of the lowest natural orbital
during the expansion is depicted in Fig.\ \ref{NatOrbN0.5}(b).  One 
can see that it exhibits exactly the same features observed in 
Refs.\ \cite{rigol04c,rigol05c} when there was no additional periodic 
potential. After the Mott insulator melts, the shape of the lobes 
of the natural orbital stops changing and they just move with the 
maximum velocity in the lattice given by Eq.\ (\ref{velocity}) for $k=k_m$. 

\begin{figure}[h]
\begin{center}
\includegraphics[width=0.5\textwidth]{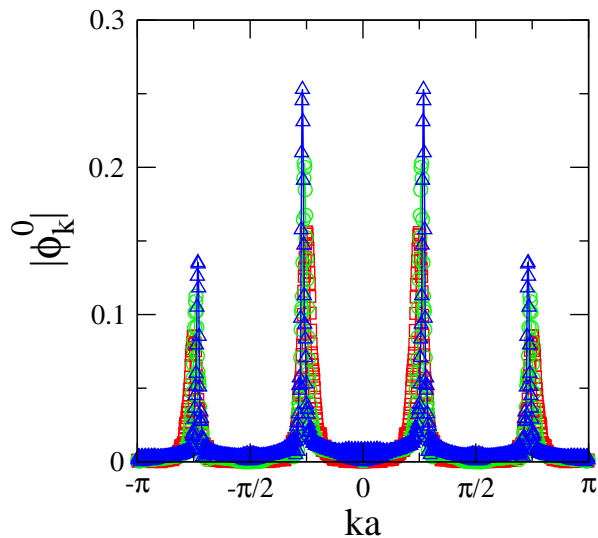}
\caption{Evolution of the Fourier transform of the lowest 
natural orbital wave-function of 100 HCB's, initially prepared 
in an insulating state with mean density of 0.5 on 1000 lattice sites. 
The superlattice parameters are $\ell=2$ and $A=2t$. The times are 
$50\hbar/t$ (\textcolor{red}{$\square$}), 
$200\hbar/t$ (\textcolor{green}{$\bigcirc$}), and 
$400\hbar/t$ (\textcolor{blue}{$\triangle$}).
}
\label{FourierN0.5}
\end{center}
\end{figure}

One final remark on these emerging quasi-condensates is in order. 
In contrast to those emerging in a system without a superlattice potential 
which are mainly formed by particles with $ka=\pm\pi/2$, we find that 
in the superlattice the quasi-condensates are mainly formed by HCB's
with the four momenta $k_m$. This is reflected by the four-peak structure 
of the Fourier transform of $\phi^\eta$ at momenta $k_m$,
depicted in Fig.\ \ref{FourierN0.5}. Hence, the four peaks that appear 
in the momentum distribution in Fig.\ \ref{PerfilesN0.5}(b)
reflect the formation of the quasi-condensates in Fig.\ \ref{NatOrbN0.5}(b), 
which have a richer structure in $k$-space than those that emerge in the 
absence of the superlattice.

\section{Conclusions}
\label{sec:conclusion}

In this work, we have presented non-trivial generalizations of a
nonequilibrium 
phenomenon originally found in systems of hard-core bosons that
expand out of strongly correlated Mott-insulating states.
First, we have shown that the emergence of coherent matter waves at finite
momenta persists away from the hard-core limit, i.e., to finite
values of the on-site repulsion $U$ which range down to the critical
values, as long as a Mott-insulating state is attainable for the
initial state.
The accurate treatment of this complicated many-body problem has been
made possible by the advent of the t-DMRG.
By comparing to exact results in the hard-core case, we have shown
that our t-DMRG calculations, although 
approximate, are very reliable. 
Two new features appear in the 
soft-core case: 
{\it (i)} Although the time evolution of
the density profiles is indistinguishable from those of hard-core
bosons at large values of $U$, 
the initial momentum distribution functions are markedly different. 
Nevertheless, after the Mott region melts, the momentum distribution
functions  of both systems become nearly identical. 
{\it (ii)} As the strength of the  
interaction is reduced, a shift of the momentum of the coherent matter wave
to values smaller than $\pi/2a$ is observed. 
The appearance of a power law in the spatial decay of the one-particle
density matrix demonstrates explicitly the (quasi-)coherent nature of the
resulting matter wave. 
Furthermore, we have shown that a coherent matter wave can be also
obtained in the presence of superlattice potentials given an adequate
selection of the initial insulating state. 
In the latter case, the emerging quasi-condensates 
have a more complicated structure in momentum space, with a number of 
dominating momenta whose locations depend on the superlattice used.  

The results of this work show that it is possible to engineer atom lasers
with a high degree of control. The momentum of the coherent matter wave
can first be regulated by setting the wavelength of the underlying
optical lattice and further
fine-tuned by regulating the depth of the potentials
in the lattice, i.e., the intensity of the corresponding laser beam. 
A further finite shift of the momentum can be achieved by superimposing
another laser beam with a commensurate wavelength, giving rise to a 
superlattice, which leads to the emergence of a matter wave with two
(or more) dominating momenta.

\ack
We are grateful to the Landesstiftung Baden-W\"urttemberg (research program
``Atomoptik''), the SFB 382, and the SFB/TR 21 for financial support. 
M.R.\ was supported by NSF-DMR-0312261 and NSF-DMR-0240918.
We thank HLR-Stuttgart (Project DynMet) and NIC-J\"ulich for allocation
of computer time.

\end{document}